\documentclass[10pt,letterpaper]{article}         %
\usepackage{opex3}                                %
\usepackage{graphicx}
\usepackage{epsfig}
\usepackage{cite}
\usepackage{color}
\usepackage{amsmath,amssymb,subfigure}
\def\U#1{{\rm #1}} 
\def\u#1{_{\rm #1}}

\newcommand{\expect}[1]{\langle #1 \rangle}

\def\Dt{\U{D}\u{T}}

\def\BGt{\U{BG}\u{T}}

\begin{document}
\title{A low-noise frequency down-conversion to the telecommunication band 
for a quantum communication based on NV centers in diamond}

\author{
Rikizo~Ikuta$^{1,*}$, 
Toshiki~Kobayashi$^{1}$,
Shuto~Yasui$^{1}$,
Shigehito~Miki$^{2}$,
Taro~Yamashita$^{2}$,
Hirotaka~Terai$^{2}$,
Mikio~Fujiwara$^{3}$,
Takashi~Yamamoto$^{1}$,
Masato~Koashi$^{3}$,
Masahide~Sasaki$^{4}$,
Zhen~Wang$^{2,5}$,
Nobuyuki~Imoto$^{1}$}
\address{
$^1$Graduate School of Engineering Science, Osaka University,
Toyonaka, Osaka 560-8531, Japan\\
$^2$Advanced ICT Research Institute, 
National Institute of Information and Communications Technology (NICT),
Kobe 651-2492, Japan\\
$^3$Photon Science Center, 
The University of Tokyo, Bunkyo-ku, Tokyo 113-8656, Japan\\
$^4$Advanced ICT Research Institute, 
National Institute of Information and Communications Technology (NICT),
Koganei, Tokyo 184-8795, Japan\\
$^5$Shanghai Center for SuperConductivity, Shanghai Institute of Microsystem and
Information Technology, Chinese Academy of Sciences, R3317, 865 Changning
Road, Shanghai 200050, PR China
}

\email{$^*$ikuta@mp.es.osaka-u.ac.jp}

\begin{abstract}
We demonstrate a low-noise frequency down-conversion of photons 
at 637 nm to the telecommunication band at 1587 nm 
by the difference frequency generation in a periodically-poled lithium niobate. 
An internal conversion efficiency of the converter is estimated to be 
0.44 at the maximum which is achieved by a pump power of 430 mW, 
whereas a rate of internal background photons caused by the strong cw pump laser 
is estimated to be 9 kHz/mW within a bandwidth of about 1 nm. 
By using the experimental values related to the intrinsic property of the converter, 
and using the intensity correlation and the average photon number of 
a 637-nm input light pulse, 
we derive the intensity correlation of a converted telecom light pulse. 
Then we discuss feasibility of a single-photon 
frequency conversion to the telecommunication band 
for a long-distance quantum communication based on NV centers in diamond. 
\end{abstract}

\ocis{
(270.5565) Quantum communications; 
(270.5585) Quantum information and processing; 
(130.7405) Wavelength conversion devices;
(190.4223) Nonlinear wave mixing
} 

\section{Introduction}
Quantum repeaters for efficiently transmitting a quantum state 
through lossy and noisy quantum channels 
have been proposed~\cite{Sangouard2011} 
and actively studied 
for realizing long distance quantum communication\cite{Gisin2007}. 
In the quantum repeater protocols, 
photons entangled with quantum memories at remote parties 
are sent to relay points, 
and they are measured by the Bell measurement 
for establishing the entanglement 
between the quantum memories at the remote parties. 
The elementary part that creates entanglement between a quantum memory 
and a photon has been demonstrated 
in atomic systems~\cite{Matsukevich2004,Ritter2012}, 
trapped ion systems~\cite{Olmschenk2009} and 
solid-state systems~\cite{Togan2010, Gao2012}. 
In such systems, 
the wavelengths of the photons are strictly limited 
by the structure of their energy levels, 
which lie around visible range.
On the other hand, 
when we look at optical-fiber networks, 
the photons at the telecommunication wavelengths 
are crucial for the efficient transmission of the photons. 
Thus a photonic quantum interface~\cite{Kumar1990} 
for visible-to-telecommunication wavelength conversion 
is considered to be inevitable 
for a long distance quantum communication 
based on the quantum repeaters~\cite{Shahriar2012}. 

So far, 
in order to build such a quantum interface, the frequency down-conversion 
working at a single-photon level 
from visible to the telecommunication wavelengths has been actively studied 
by using nonlinear optical media~\cite{Radnaev2010, Dudin2010,
Takesue2010, Curtz2010, Zaske2011, Ikuta2011, Zaske2012, Ikuta2013}. 
In the first experimental demonstration 
employing a light field with a non-classical property, 
the wavelength of 795 nm which corresponds to the D1 line 
of the rubidium atoms are converted to 1367 nm 
by using a third-order optical nonlinearity 
in a rubidium atomic cloud~\cite{Radnaev2010, Dudin2010}. 
In Ref.~\cite{Dudin2010}, 
preservation of the entanglement between two photons 
after the frequency down-conversion was also observed. 
For a wide-band and a compact wavelength conversion, 
the difference frequency generation~(DFG) 
via a second-order optical nonlinearity 
in a periodically-poled lithium niobate~(PPLN) has been used 
and the preservation of the quantum state 
through the wavelength conversion from 780 nm 
which corresponds to the D2 line of the rubidium atom 
to 1522 nm has been demonstrated~\cite{Ikuta2011, Ikuta2013}. 
The non-classical property of a light converted to 1313 nm 
from a 711-nm light emitted by a quantum dot 
has also been demonstrated~\cite{Zaske2012}. 
In Ref.~\cite{Zaske2011}, 
the feasibility of the frequency down-conversion 
of a light at a single-photon level 
from 738 nm to 1557 nm by using the 1403-nm pump light 
has been shown for single-photon sources 
based on silicon-vacancy centers in diamond~\cite{Neu2011}. 

Another important wavelength 
which should be converted to the telecommunication bands 
for quantum communications is 637 nm corresponding to 
the resonant wavelength of the nitrogen-vacancy~(NV) center 
in diamond~\cite{Doherty2013}. 
The NV center in diamond is 
one of the promising candidates for the quantum memories 
with the ability to create the entanglement with photons~\cite{Togan2010}. 
In spite of a variety of demonstrations 
related to the NV centers in diamond 
in recent years~\cite{Batalov2008,Schoroder2011,Hausmann2011,Mizuochi2012,Schroder2012,Bernien2013, Albrecht2013}, 
to the best of our knowledge, 
none of the demonstration of the frequency down-conversion 
from 637 nm to the telecommunication bands 
with the preservation of the non-classical property and/or 
the quantum information of the input light 
has been performed. 
In the experiment in Ref.~\cite{Pelc2010}, 
while the conversion from 633 nm to 1.56-$\mu$m 
has been observed with a 1.06-$\mu$m pump light, 
the signal-to-noise ratio~(SNR) was not sufficient 
for the observation of the above mentioned properties. 

In this paper, 
we demonstrate a low-noise wavelength conversion of a 637-nm light 
at a single-photon level to the telecommunication band 
by using a PPLN waveguide 
for realizing the efficient optical-fiber quantum communication 
based on NV centers in diamond~\cite{Shahriar2012}. 
The converted wavelength in this experiment is 1587 nm 
which is determined by the wavelength of 1064 nm of the cw pump light. 
In the experiment, we evaluate the performance of the wavelength converter, 
namely we clarify the overall conversion efficiency of the signal light 
and the amount of background photons induced by the conversion process. 
From the experimental results, 
we derive 
the intensity correlation and the average photon number 
of a converted telecom light pulse 
which are decided by 
those of a 637-nm input light pulse 
and the intrinsic property of the converter. 
Then we discuss the feasibility of the observation 
of non-classical property of the telecom light 
after the frequency down-conversion from a 637-nm light pulse. 
We found that 
when the transform-limited light pulse from the NV centers in diamond 
with a lifetime of 13.7 ns reported in Refs.~\cite{Hausmann2011, Mizuochi2012}
is prepared with a probability above $10^{-3}$, 
the intensity correlation of the converted telecom photon 
will be below $0.75$ which indicates 
the non-classical property of the converted light. 

\section{Experiment}
\subsection{Experimental setup}
\begin{figure}[t]
 \begin{center}
 \scalebox{0.6}{\includegraphics{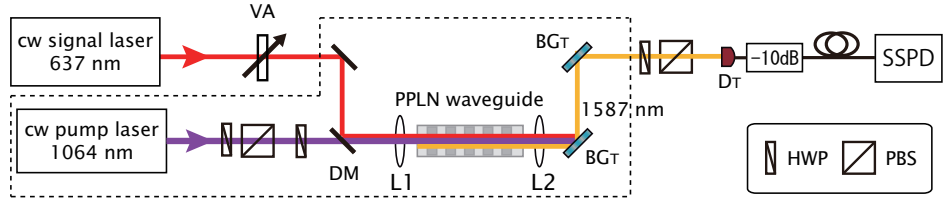}}
  \caption{ Experimental setup. 
  The cw signal light at 637 nm is frequency down-converted 
  to 1587 nm by using a strong cw pump light at 1064 nm. 
  The converted light is diffracted 
  by the two Bragg gratings~($\BGt$) 
  and detected by the SSPD. 
  \label{fig:setup}}
 \end{center}
\end{figure}
The experimental setup is shown in Fig.~\ref{fig:setup}. 
The 637-nm signal light 
is taken from a vertical~(V-) polarized cw external cavity laser diode 
with the linewidth of $< 150$ MHz and a maximum power of 9 mW. 
The power of the signal light input to the PPLN waveguide 
is adjusted by using a variable attenuator~(VA) 
within the range from $\approx 1$mW 
to a photon rate of $\approx 40$ MHz. 
We use a cw pump laser 
with the center wavelength of 1064 nm and the linewidth of $<100$ MHz 
for the wavelength conversion. 
The power of the pump laser coupled to the PPLN waveguide 
is varied up to $\approx 400$ mW 
by a polarization beamsplitter~(PBS) and a half-wave plate~(HWP). 
The pump beam is set to the vertical polarization by a HWP and then 
it is combined with the signal beam by a dichroic mirror. 
The two beams are coupled to the PPLN waveguide 
by using an aspheric lens~(L1) of $f=8$mm 
with anti-reflective coating for both the signal light and the pump light. 

The PPLN waveguide we used is a ridged-type waveguide, 
the cross section is square with 8-$\mu$m wide, 
and the length of the waveguide is 23 mm. 
The period of periodically-poled structure is 11.5 $\mu$m~\cite{Nishikawa2009}. 
Due to the type-0 quasi-phase matching condition, 
the V-polarized signal photon at 637 nm is converted 
to the V-polarized telecom photon at 1587 nm 
with the V-polarized strong pump beam at 1064 nm. 
The temperature is controlled to be $\approx 20^\circ$C. 
The acceptable bandwidth for 637-nm signal light of this waveguide 
is calculated to be $\approx$ 0.1 nm. 

The output photons from the PPLN waveguide are collimated 
by an aspheric lens~(L2) of $f=8$mm 
with anti-reflective coating for the signal light, 
pump light and telecom light. 
Two Bragg gratings~($\BGt$) with a bandwidth of 1 nm 
extract only the converted telecom light among them. 
The combined bandwidth of two $\BGt$ is $\tilde{\Delta}\approx 0.7$ nm. 
The telecom light is coupled to the single-mode fiber~($\Dt$), 
and detected by a superconducting single-photon detector~(SSPD) 
after passing through an attenuator of $\approx$ 10 dB 
for preventing the saturation of the detection counts. 
The SSPD has a cavity structure and 
its quantum efficiency denoted by $\epsilon$ is $\approx$ 60\%~\cite{Miki2013}. 

\subsection{Experimental results}
In order to estimate an internal conversion efficiency 
of the frequency down-converter 
and an amount of background photons generated by the conversion process, 
we first estimate the transmittance of the overall optical circuit 
by measuring the power of the light before and after the optical components. 
The coupling efficiency of the signal light 
to the PPLN waveguide is $T\u{in}\approx 0.88$. 
The diffraction efficiency of each of $\BGt$ is $T\u{BG}\approx 0.86$. 
The telecom photons diffracted by the last $\BGt$ 
pass through the HWP and PBS, and then enter $\Dt$. 
The efficiency of these processes is estimated to be $\approx 0.42$. 
As a result, the transmittance of the optical circuit 
from the first $\BGt$ to the front of the fiber attenuator 
is $T\u{oc}\approx 0.31$. 
The transmittance of the fiber attenuator 
is $T\u{att}\approx 0.083$. 

\begin{figure}[t]
 \begin{center}
 \scalebox{0.55}{\includegraphics{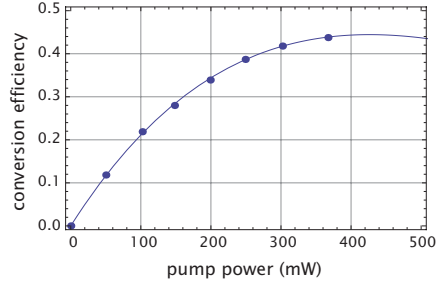}}
  \caption{
  (a)
  Conversion efficiency 
  from 637 nm to 1587 nm 
  as a function of the pump power $P$. 
  The curve fitted to the experimental data is described 
  by $T\u{conv}^{\U{max}}\sin^2(\sqrt{\eta P})$ 
  with $T\u{conv}^{\U{max}}\approx 0.44$ and $\eta \approx 5.8/\U{W}$. 
  \label{fig:conv}}
 \end{center}
\end{figure}
The internal conversion efficiency in the PPLN 
depending on the pump power is estimated 
by the observed output power of the converted light coupled to $\Dt$ 
connected to a power meter 
instead of the fiber attenuator and the SSPD in Fig.~\ref{fig:setup}. 
In this measurement, 
we set the power of the input light at 637 nm 
coupled to the PPLN waveguide to be $\approx 840\mu$W. 
From the observed values by the power meter 
and the estimated transmittance $T\u{oc}$ of the optical components, 
we estimate the internal conversion efficiencies of the photon number. 
The result is shown in Fig.~\ref{fig:conv}. 
The conversion efficiency of the DFG process with a strong pump light 
is expected to be proportional to 
$\sin^2(\sqrt{\eta P})$~\cite{Kumar1990}, 
where $P$ represents the pump power. 
The best fit to the observed conversion efficiency 
with $T\u{conv}=T\u{conv}^{\U{max}}\sin^2(\sqrt{\eta P})$ 
gives $T\u{conv}^{\U{max}}\approx 0.44$ and $\eta \approx 5.8/\U{W}$. 
The best conversion efficiency of 0.44 will be achieved 
at a pump power of $\approx 430$\,mW. 

\begin{figure}[t]
 \begin{center}
 \scalebox{0.55}{\includegraphics{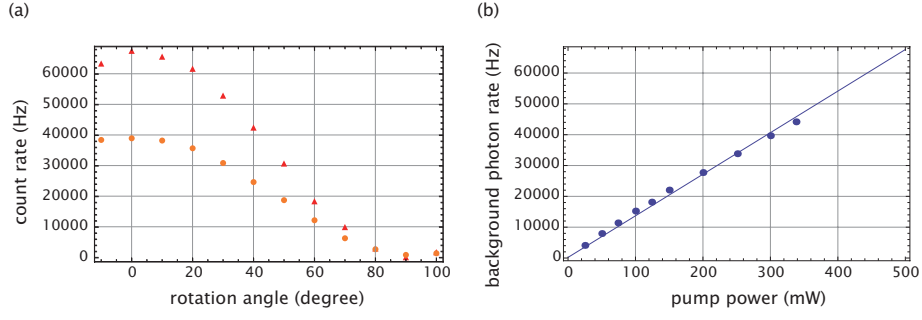}}
  \caption{
  (a)
  The observed count rates of the background photons~(orange circle) 
  and the signal photons~(red triangle) depending on 
  the rotation angle of the HWP followed by the PBS and the detector. 
  The degrees of polarization of the background photons and the signal photons 
  were 0.95 and over 0.99, respectively. 
  The result shows that the background photons include the H polarization. 
  (b)
  The dependency of the background photon rate on the pump power. 
  By using the dark count rate of $d = 220$ Hz, 
  the background photon rate is fitted to the 
  the line described by $B P+d$ with $B \approx 135$ Hz/mW. 
  \label{fig:noise}}
 \end{center}
\end{figure}
We evaluate 
the amount of background photons at the down-converted wavelength of 1587 nm. 
The background photons were measured 
when we turned off the signal light and 
input only the pump light to the PPLN waveguide. 
Before we investigate the dependency of the background photons on the pump power, 
we first see the polarization property of the background photons, 
which was performed by measuring the background photon counts 
with various rotation angles of the HWP in front of the last PBS. 
This measurement was performed at 300-mW pump power. 
The experimental result of the count rate of the background photons 
depending on the rotation angle is shown in Fig.~\ref{fig:noise}~(a). 
The degree of polarization of the background photons was $\approx 0.95$. 
The standard deviations of the observed values 
under the assumption of the Poisson statistics of the observed counts 
are less than $0.01$. 
We note that when we measured the polarization property for the signal light, 
the degree of the polarization of the signal photons was over $0.99$ 
by subtracting the effect of the background photons from the observed counts. 
From the result, we see that while the polarization of the background photons 
are almost V polarized, a small number of the background photons are 
orthogonal to V polarization. 
In the following experiment, 
H polarization component of the background photons 
is reflected by the PBS in front of $\Dt$. 

The dependency of the background photon rate on the pump power 
is shown in Fig.~\ref{fig:noise}~(b). 
We see that the background photon rate linearly increase with the pump power. 
From this, we infer that the main cause of the noises is 
the Raman scattering of the strong pump light 
as reported in the previous experiments 
of the frequency down-conversion~\cite{Zaske2011,Ikuta2013}. 
In order to fit the data, 
we use the linear function of the photon count rate written as $N=B P+d$, 
where $d=220$Hz is determined by an observed value of the dark count rate of the SSPD. 
The coefficient $B$ is estimated to be 135 Hz/mW. 
The internal background photon rate of the converter is 
$B/(T\u{oc}T\u{att}\epsilon)\approx 9\times 10^3$ Hz/mW. 
As a reference, we compare 
the intrinsic background photon generation rate due to the Raman scattering 
in this experiment 
with that in the demonstration for frequency down-conversion 
from 780 nm to 1522 nm in Refs.~\cite{Ikuta2011, Ikuta2013} 
when these converters give the same internal conversion efficiency. 
The observed value of the Raman scattering in Ref.~\cite{Ikuta2013} 
was $B_{780}=80$Hz/mW which was detected 
by a photon detector with the quantum efficiency of 0.125. 
In the report, the overall conversion efficiency 
just before the photon detector was 0.005 
which includes the internal conversion efficiency of 0.07. 
From these values, the internal background photon rate 
is roughly estimated to be $\approx 9\times 10^3$ Hz/mW. 
The conversion efficiency for the cw signal light was 
$T\u{conv,780}=T\u{conv,780}^{\U{max}}\sin^2(\sqrt{\eta\u{780}P})$ 
with $T\u{conv,780}^{\U{max}}\approx 0.71$ 
and $\eta\u{780} \approx 3.6/\U{W}$~\cite{Ikuta2011, Ikuta2013}. 
When the conversion efficiencies of the two converters are the same, 
the ratio of the background photon rate of the converter 
for the 637-nm light 
to that for the 780-nm light takes a maximum 
for $T\u{conv}=T\u{conv,780}=T\u{conv}^\U{max}$. 
For the converter in Refs.~\cite{Ikuta2011, Ikuta2013}, 
$T\u{conv,780}=T\u{conv}^\U{max}$ 
is achieved at a pump power of $\approx 230$\,mW, 
resulting in the background photon rate of $\approx 2.1\times 10^6$\,Hz. 
For the converter in this paper, 
$T\u{conv}=T\u{conv}^\U{max}$ is given by 
the pump power of $\approx 430$\,mW, 
resulting in the background photon rate of $\approx 3.7\times 10^6$\,Hz. 
By using these values, 
the background photon rate 
introduced by the conversion process in this experiment is, 
at the most, twice as large as that in Refs.~\cite{Ikuta2011, Ikuta2013}. 

\section{Discussion}
\begin{figure}[t]
 \begin{center}
 \scalebox{0.8}{\includegraphics{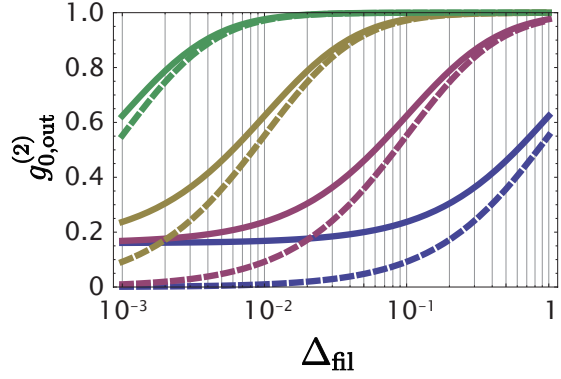}}
  \caption{
  The estimated $g^{(2)}\u{0,out}$ of the converted telecom light 
  as a function of $\Delta\u{fil}$ 
  in the cases of $g^{(2)}\u{0,in}\approx 0.16$~(solid curves) 
  and $g^{(2)}\u{0,in}\approx 0$~(dotted curves). 
  The measurement time is fixed to be $\tau\u{time} = 52\,\U{ns}$. 
  The four curves show the intensity correlation 
  in the cases of $\bar{n}\u{in,sig}=1$, $10^{-1}$, $10^{-2}$ and $10^{-3}$ 
  from the bottom. 
  \label{fig:snr0}}
 \end{center}
\end{figure}
In this section, 
we discuss the performance of 
the demonstrated frequency down-converter 
for a pseudo single photon pulse 
emitted from the NV center in diamond 
based on our experimental results in the previous section. 
Here we estimate the well-known intensity correlation $g^{(2)}_{0}$ 
of the converted light pulse to see the ability 
of the frequency down-converter 
to preserve the non-classical property $g^{(2)}_{0} < 1$. 
We regard a part of the optical circuit 
including the PPLN and the two BGs 
as an essential part of the converter, 
which is the region surrounded by dashed lines in Fig.~\ref{fig:setup}. 
In the following, we derive 
the intensity correlation and the average photon number 
of a converted telecom light pulse from the converter 
when those of a 637-nm signal light pulse are given. 

Let us first derive the intensity correlation of a light pulse X 
which is a mixture of two statistically-independent light pulses A and B 
with the intensity correlation $g^{(2)}_{0,\U{A}}$ and $g^{(2)}_{0,\U{B}}$ 
and the average photon number $\expect{\hat{n}\u{A}}$ and $\expect{\hat{n}\u{B}}$, 
where $\hat{n}\u{A(B)}$ represents the photon number operator. 
The intensity correlation of each pulse is given by 
$g^{(2)}_{0,i}=\expect{\hat{n}_i(\hat{n}_i-1)}/\expect{\hat{n}_i}^2$ 
for $i=\U{A,B,X}$. 
We also define the Fano factor 
$F_i=\expect{(\Delta \hat{n}_i)^2}/\expect{\hat{n}_i}$, 
where $\Delta \hat{n}_i=\hat{n}_i-\expect{\hat{n}_i}$. 
We then obtain $\expect{\hat{n}_i}(g^{(2)}_{0,i}-1)=F_i-1$. 
When the two light pulses A and B are statistically independent 
so that $\expect{\Delta n\u{A}\Delta n\u{B}}=0$, 
the Fano factor of the light pulse X is calculated to be 
$F\u{X}=\expect{(\Delta \hat{n}\u{A}+\Delta \hat{n}\u{B})}^2
/\expect{\hat{n}\u{A}+\hat{n}\u{B}}
=F\u{A}\expect{\hat{n}\u{A}}/\expect{\hat{n}\u{A}+\hat{n}\u{B}}
+F\u{B}\expect{\hat{n}\u{B}}/\expect{\hat{n}\u{A}+\hat{n}\u{B}}$. 
As a result, the intensity correlation of the light pulse X 
is given by 
\begin{eqnarray}
g^{(2)}_{0,\U{X}} &=& \frac{1}{\expect{\hat{n}\u{A}+\hat{n}\u{B}}^2}
\left(
\expect{\hat{n}\u{A}}^2g^{(2)}_{0,\U{A}}
+\expect{\hat{n}\u{B}}^2g^{(2)}_{0,\U{B}}
+2\expect{\hat{n}\u{A}}\expect{\hat{n}\u{B}}
\right) \\
&=&
\frac{1}{(1+\zeta)^2}
\left(
\zeta^2g^{(2)}_{0,\U{A}}
+g^{(2)}_{0,\U{B}}
+2\zeta
\right), 
\label{eq:g2}
\end{eqnarray}
where $\zeta=\expect{\hat{n}\u{A}}/\expect{\hat{n}\u{B}}$ is 
the ratio between the average photon numbers of the two light pulses. 
As we have seen in the previous section, 
the frequency down-converter adds the background photons in the output mode. 
Therefore the above relation~(\ref{eq:g2}) is useful 
for the estimation of the output light pulse. 

Suppose that 
the intensity correlation of a 637-nm input light pulse is 
$g^{(2)}_{0,\U{in}}$ and the average photon number of the light 
is $\bar{n}\u{in,sig}$. 
For simplicity, 
we assume that the spectral width of the input light 
is much narrower than the acceptable bandwidth $\approx$ 0.1 nm 
of the frequency down-converter. 
We also assume that the background photons 
induced by the pump light follow the Poisson statistics, 
and are independent of the down-converted photons. 
It is convenient to replace these background photons 
by a noise source placed at the input of the down-converter. 
The average photon number $\bar{n}\u{in,noise}$ of this equivalent input noise 
is then given by $\bar{n}\u{in,noise}
=BP\Delta\u{fil}\tau\u{time}/(T\u{in}T\u{conv}T\u{oc}T\u{att}\epsilon)$. 
Here $\tau\u{time}$ is a measurement time window, 
and $\Delta\u{fil}$ is a ratio of a filter bandwidth 
to the bandwidth of $\tilde{\Delta}\approx 0.7$ nm in the current experiment. 
Note that we assume that the bandwidth of the filter 
is much wider than the spectral width of the 637-nm signal light. 
The reason why the component of the background photons 
is proportional to $\tau\u{time}$ and $\Delta\u{fil}$ 
is that the background photons induced by the cw pump light 
for the wavelength conversion are temporally continuous and spectrally broad. 
From the observed values in the previous section as 
$B\approx 135$ Hz/mW, 
$T\u{in}\approx 0.88$, 
$T\u{oc}\approx 0.31$, 
$T\u{att}\approx 0.083$, 
$\epsilon\approx 0.6$ 
and 
$T\u{conv}=T\u{conv}^{\U{max}}\sin^2(\sqrt{\eta P})$ 
with $T\u{conv}^{\U{max}}\approx 0.44$ and $\eta \approx 5.8/\U{W}$, 
we calculate as 
$\bar{n}\u{in,noise}
\approx 2.2\times 10^4\Delta\u{fil}\tau\u{time}P/\sin^2(\sqrt{5.8 P/\U{W}})$. 
By substituting 
$(g^{(2)}_{0,\U{A}}, \expect{\hat{n}\u{A}})$ 
and 
$(g^{(2)}_{0,\U{B}}, \expect{\hat{n}\u{B}})$ 
in Eq.~(\ref{eq:g2}) 
by 
$(g^{(2)}_{0,\U{in}}, \bar{n}\u{in,sig})$ and 
$(1, \bar{n}\u{in,noise})$, respectively, 
and by using $\zeta\u{in}=\bar{n}\u{in,sig}/\bar{n}\u{in,noise}$, 
the intensity correlation $g^{(2)}_{0,\U{out}}$ of the output light pulse 
is given by 
\begin{eqnarray}
g^{(2)}_{0,\U{out}} = \frac{1}{(1+\zeta\u{in})^2}
\left(
\zeta\u{in}^2
g^{(2)}_{0,\U{in}}
+1
+2\zeta\u{in}
\right). 
\label{eq:g2out}
\end{eqnarray}
Because $T\u{in}T\u{conv}T\u{BG}^2\approx 0.3\sin^2(\sqrt{5.8P/\U{W}})$ is 
the transmittance of the intrinsic part of the converter, 
we obtain the average photon number $\bar{n}\u{out}$ of the output light pulse 
to be 
\begin{eqnarray}
\bar{n}\u{out}
=T\u{in}T\u{conv}T\u{BG}^2 (\bar{n}\u{in,sig}+\bar{n}\u{in,noise}). 
\end{eqnarray}

Below we borrow the observed value of $g^{(2)}\u{0,in}\approx 0.16$ and 
a lifetime of $\tau \approx 13.7$ ns for the 637-nm light pulse 
from the NV center in the diamond nanocrystal reported in Ref.~\cite{Hausmann2011}, 
and then we estimate the expected value of $g^{(2)}\u{0,out}$ 
for the converted telecom light. 
We assume that the spectral width of the 637-nm light is 
close to the transform-limit so that it is much narrower than 
$\Delta\u{fil}\tilde{\Delta}$. 
We choose $\tau\u{time}=52\,\U{ns}$ 
to collect the pulsed signal light over 99\%, 
and use $P\approx 430$ mW 
which achieves the maximum conversion efficiency in our experiment. 
From Eq.~(\ref{eq:g2out}) and $g^{(2)}\u{0,in}\approx 0.16$, 
we can calculate $g^{(2)}\u{0,out}$ 
which is shown as a function of $\Delta\u{fil}$ 
for $\bar{n}\u{in,sig}=10^{-3}, 10^{-2}, 10^{-1}, 1$ 
in Fig.~\ref{fig:snr0}. 
As a reference, 
we also show $g^{(2)}\u{0,out}$ for the converted telecom light 
when the 637-nm signal light has $g^{(2)}\u{0,in}=0$ and the same lifetime. 

\begin{figure}[t]
 \begin{center}
 \scalebox{0.6}{\includegraphics{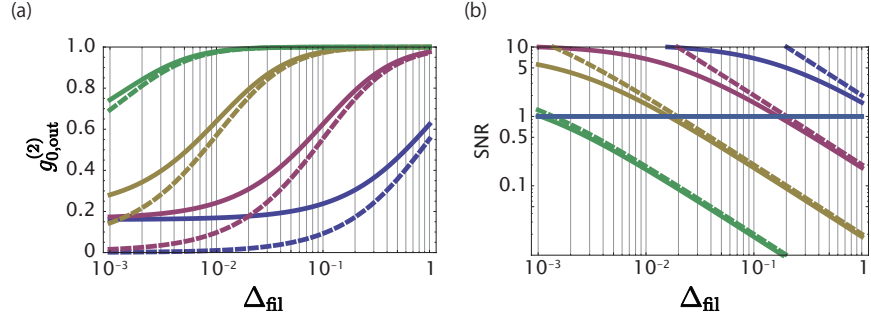}}
  \caption{
  The expected observed values of $g^{(2)}\u{0,out}$ and SNR of 
  the converted telecom light 
  as a function of $\Delta\u{fil}$ 
  in the cases of $g^{(2)}\u{0,in}\approx 0.16$~(solid curves) 
  and $g^{(2)}\u{0,in}\approx 0$~(dotted curves). 
  The measurement time is fixed to be $\tau\u{time} = 52\,\U{ns}$. 
  (a) The four curves show the intensity correlation 
  in the cases of $\bar{n}\u{in,sig}=1$, $10^{-1}$, $10^{-2}$ and $10^{-3}$ 
  from the bottom. 
  (b) The four curves show the SNR 
  in the cases of $\bar{n}\u{in,sig}=1$, $10^{-1}$, $10^{-2}$ and $10^{-3}$ 
  from the top. 
  \label{fig:snr}}
 \end{center}
\end{figure}
In practice, the apparent intensity correlation of the light 
measured by the Hanbury-Brown and Twiss setup~\cite{Hanbury1956} 
is affected by the dark count of the detectors. 
We assume that coupling efficiencies to the single-mode fibers and 
two photon detectors for the Hanbury-Brown and Twiss setup 
are the same as those used in our experiment, namely 
the transmittance of the optical circuit after the PPLN 
is $T\u{oc}\approx 0.31$, 
the quantum efficiency is $\epsilon\approx 0.6$, 
and the dark count rate is $d = 220$ Hz. 
When a half BS is used for the measurement of $g^{(2)}_{0,\U{out}}$, 
an expected observed value of $g^{(2)}_{0,\U{out}}$ 
for the converted telecom light as a function of $\Delta\u{fil}$ 
is shown in Fig.~\ref{fig:snr}~(a). 
From the figure, 
even for $\bar{n}\u{in,sig}=10^{-3}$, 
we see a non-classical intensity correlation 
$g^{(2)}\u{0,out}\approx 0.74$ of the converted telecom light 
at $\Delta\u{fil}=1.2\times 10^{-3}$ which corresponds to about $100$ MHz. 
Using Eq.~(\ref{eq:g2}), 
we also calculate an expected SNR of the converted telecom light at the each detector 
under the assumption that the 637-nm light is a mixture 
of two statistically-independent light sources 
composed of an ideal single photon and 
background photons that follow the Poisson statistics, 
and show the results in Fig.~\ref{fig:snr}~(b). 

In the above estimation, 
we assumed the spectral bandwidth of the 637-nm signal light 
is much smaller than the acceptable bandwidth $\approx$ 0.1 nm 
of PPLN crystal and the filter bandwidth $\Delta\u{fil}\tilde{\Delta}$, 
where $\tilde{\Delta}\approx 0.7$ nm, 
while it was not satisfied in the experiment reported in Ref.~\cite{Hausmann2011} 
due to the inhomogeneous broadening at the room temperature. 
On the other hand, 
recently a transform-limited narrow bandwidth light pulse 
from the zero phonon line of a NV center 
in a bulk diamond at a low temperature has been reported in Ref.~\cite{Batalov2008}. 
In addition, there are several studies for efficient collection of the light from
NV center using a solid immersion lens~\cite{Schoroder2011} 
and a photonic crystal cavity~\cite{Englund2010}. 
When such experimental efforts will 
meet our requirements of the 637-nm signal light in the near future, 
the demonstrated low noise frequency down converter 
will achieve the conversion of the non-classical light 
from the NV centers to the telecom band. 

\section{Conclusion}
In conclusion, we have demonstrated 
that the low noise frequency down-conversion of 
the photon at 637 nm which corresponds to a resonant wavelength 
of the NV center in diamond to the telecommunication band. 
The analysis based on our experimental results shows that 
if the transform-limited light pulse from the NV center in diamond 
is prepared even with a probability of $10^{-3}$, 
the intensity correlation of the telecom light 
after the frequency down-conversion 
with the proper frequency filtering will be well below 1. 
We believe that the result will lead to the efficient connection 
between solid-state devices composed of the NV center in diamond 
and telecom fiber networks. 

\section*{Acknowledgements}
This work was supported by the Funding Program 
for World-Leading Innovative R \& D 
on Science and Technology~(FIRST), 
MEXT Grant-in-Aid for Scientific Research 
on Innovative Areas 21102008, 
MEXT Grant-in-Aid for Young scientists(A) 23684035, 
JSPS Grant-in-Aid for Scientific Research(A) 25247068 and (B) 25286077. 

\end{document}